# Interlayer Coupling and the Metal-Insulator Transition in Pr-substituted $Bi_2Sr_2CaCu_2O_{8+y}$

*C. Quitmann, B. Beschoten\*, R. J. Kelley, G. Güntherodt\*, M. Onellion*

*Department of Physics, University of Wisconsin, 1150 University Ave., Madison, WI 53706, USA*

*\*2. Physikalisches Institut, RWTH-Aachen, D-52064 Aachen, Germany*

Dr. Christoph Quitmann
Department of Physics
University of Wisconsin
475 N. Charter St, B630-A
**Madison, WI 53706**
    U S A
phone: 608-262-9248
fax:    608-265 2334
Email: **quitmann@macc.wisc.edu**

# Interlayer Coupling and the Metal-Insulator Transition in Pr-substituted $Bi_2Sr_2CaCu_2O_{8+y}$

C. Quitmann, B. Beschoten*, R. J. Kelley, G. Güntherodt*, M. Onellion
Department of Physics, University of Wisconsin, 1150 University Ave., Madison, WI 53706, USA
*2. Physikalisches Institut, RWTH-Aachen, D-52056 Aachen, Germany

Substitution of rare earth ions for Ca in $Bi_2Sr_2CaCu_2O_{8+y}$ is known to cause a metal-insulator transition. Using resonant photoemission we study how this chemical substitution affects the electronic structure of the material. For the partial Cu-density of states at $E_F$ and in the region of the valence band we observe no significant difference between a pure superconducting and an insulating sample with 60% Pr for Ca substitution. This suggests that the states responsible for superconducting are predominately O-states. The Pr-4f partial density of states was extracted utilizing the Super-Coster-Kronig Pr 4d-4f resonance. It consists of a single peak at 1.36 eV binding energy. The peak shows a strongly asymmetric Doniach-Šunjic lineshape indicating the presence of a band of electronic states with a cut off at $E_F$ even in this insulating sample. This finding excludes a band gap in the insulating sample and supports the existence of a mobility gap caused by spatial localization of the carriers. The presence of such carriers at the Pr-site between the $CuO_2$-planes shows that the electronic structure is not purely 2-dimensional but that there is finite interlayer coupling. The resonance enhancement of the photoemission cross section at the Pr-4d threshold was studied for the Pr-4f and for Cu-states. Both the Pr-4f and the Cu-states show a Fano-like resonance. This resonance of Cu-states with Pr-states is another indication of coupling between the Pr-states and those in the $CuO_2$-plane. Because of the statistical distribution of the Pr-ions this coupling leads to a non periodic potential for the states in the $CuO_2$-planes which can lead to localization and thus to the observed metal-insulator transition.

**PACS-Numbers :** 74.62.Dh, 74.25.Jb, 74.20.Mn

**Introduction**

Despite tremendous effort even the normal state of high temperature superconductors (HTSC) is still not understood. The so called "metallic" normal state evolves from doping a charge transfer insulator with cations of a different valence. In the case of the $Bi_2Sr_2[Ca_{1-x}RE_x]Cu_2O_{8+y}$ compound (Bi-2212) the insulating parent compound contains only RE ions and no Ca (x=1), whereas the pure HTSC contains no RE-ions but only Ca (x=0)[1]. Because Ca has a lower valence than the RE-ions, 2+ vs. 3+, it acts as an acceptor providing hole like carriers[2]. These are transferred to the CuO2-plane by a mechanism which is not understood up to now. For sufficiently high carrier density the material becomes metallic and superconducting.

The question we will address in this paper is to what extend the presence of the **RE**-ions affects the electronic states in the $CuO_2$-plane which are believed to dominate the transport properties. For our study we have used Pr as a **RE**-ion. The layered, modular structure of these materials has lead to the approach of viewing the (**RE**/Ca)-O layer only as a charge reservoir, which becomes irrelevant after donating holes to the $CuO_2$-plane. We will show that this simple concept is not correct, because there is a non negligible coupling between electronic states on the Pr-ion and those in the $CuO_2$-plane. A further question which we address is the long-standing controversy whether or not there is a genuine band gap in the insulating materials. From studying the line shape of the Pr-4f core level we conclude that this is not the case. We discuss our results in terms of a disorder induced metal-insulator transition caused by the statistical distribution of Pr-ions in this material.

**Experimental**

All measurements were performed on single phase, polycrystalline $Bi_2Sr_2[Ca_{1-x},\mathbf{Pr_x}]Cu_2O_{8+y}$ (Bi-2212) samples[3]. The sample quality was checked by X-ray diffraction. All reflexes could be indexed using the orthorhombic (*A 2 a a*) space group[4]. The samples were prepared in air and are thus overdoped.

Polycrystalline samples were chosen because in this case the photoemission spectra give the angle integrated electronic density of states N(E). Photoemission experiments were performed on the 6m-TGM monochromator at the Synchrotron Radiation Center in Stoughton, Wisconsin with photon energies in the region 20 eV $\leq$ h$\nu$ $\leq$ 140eV. Samples were transferred into the UHV-chamber using a load lock. This eliminates the need to bake the samples and preserves their original oxygen content. The sample surface was scraped in-situ to obtain clean surfaces. During the experiments the pressure was always kept below $2 \cdot 10^{-10}$ Torr. The samples were checked repeatedly for surface contamination using spectra with a photon energy of h$\nu$ = 77 eV as a reference. In case of detectable contamination the surface was scraped again. All experiments were performed at an incidence angle of 45 degrees. Photoelectrons were collected at normal emission using a 50 mm hemispherical analyzer (Vacuum Sciences Workshop). The combined resolution of the monochromator and the analyzer was typically 0.25 eV. The photon flux of the monochromator was measured using a Au-diode. The diode current was corrected for photoyield effects and these data were used to normalize the spectra. All spectra shown are raw data normalized to the photon flux of the monochromator. No background from secondary electrons was subtracted. All measurements were done at 40K, where the pure material is superconducting but the Pr-doped sample is insulating (d$\rho$/dT < 0).

## Metal-Insulator Transition

Figure **1** shows the electrical resistivity $\rho$(T) for a series of Pr-substituted Bi-2212 samples. The samples span the entire region from the pure HTSC (x=0) through the metal-insulator transition at $x_C$ =(0.49$\pm$.02)[3] up to insulating samples with x=0.8. The critical concentration is defined as the Pr-content at which the resistivity at low temperatures changes to insulating behavior (d$\rho$/dT<0). The critical concentration of $x_c$ = (0.49$\pm$.02) for Pr-substitution is in very close to that observed for doping with other **RE**-ions (Y: 0.57, Nd: 0.47, Gd: 0.49)[3] which are known to be trivalent. For tetravalent **RE**-ions on the other hand one observes a much lower critical concentration (Ce: $x_C$ = 0.23)[5]. This indicates a Pr-valence close to 3+. Such a conclusion is supported by the magnetic susceptibility which shows a Curie-law with an effective moment of $\mu_{eff}$ = 3.4 $\mu_B$ [3,6,7] close

to the value of 3.58 $\mu_B$ expected for a trivalent Pr[4f $^2$] configuration and by Hall effect measurements[3]. The data of Fig. **1** are in good agreement with those found by other research groups [8,9]. The photoemission spectra, which will be reported in the following sections, were taken on pure samples (x = 0) and insulating samples with (x = 0.6). The resistivity of both samples is shown in Fig. **1**. The sample with x = 0.6 has a clear upturn of the resistivity at low temperatures. Such an upturn (d$\rho$/dT < 0) is characteristic of an insulator. It indicates that conductivity takes place either by carriers which are thermally excited across a band gap, or by thermally activated hopping between spatially localized states. The later case, called variable range hopping, is often observed in disordered media[10]. Based on the temperature dependence of the resistivity many authors have suggested that such variable range hopping occurs in HTSC's[11, 12, 13]. In the following, we present supporting spectroscopic evidence that the metal-insulator transition is not caused by the opening of a band gap, but by spatial localization.

The photoemission spectra reported below were taken at T=40K, for which the pure sample is in the superconducting state. The sample with 60% Pr for Ca-substitution, on the other hand, is clearly in the insulating regime. Its resistivity is about 40 m$\Omega$cm, which is a factor of 30 higher than the resistivity of the pure sample even at room temperature. In the following section we will show what effect the metal-insulator transition has on the Cu-derived electronic states in these materials.

### Partial Cu-density of states

In a simple picture for the photoemission process, the incoming photon transfers its energy to a single electron in the solid, thereby removing the electron from the solid. This outgoing electron can then be detected. In this approximation the photoemission intensity is proportional to the density of electronic states. The energy dependence for the photoemission cross section $\sigma(E)$ is given by the structure of the atomic orbitals[14]. This energy dependence is usually weak and monotonic. In actual solids, on the other hand, resonance effects are known to exist. This means that above a threshold the photoemission cross section for transitions from a single angular momentum state $\ell$ of a specific element is significantly enhanced. The threshold is given by the low lying, quasi-atomic, electronic levels and thus largely independent of the chemical environment. For photonenergies significantly higher than the threshold the enhancement disappears and the photoemission cross section reduces to the value before the resonance enhancement. Because of the significant enhancement of a particular transition, such resonances are used to extract the so called partial density of states for a single element and angular momentum state. A more detailed discussion of this effect will be given in below. In the following resonant photoemission will be used to extract the partial Cu-3d density of states.[15]

We have measured the energy distribution curves (EDC) of the pure Bi-2212 material in the vicinity of the Cu-3p core level at $E_{Cu-3p} = 74.5$ eV. These spectra are shown for different photon energies in Fig. **2**. For comparison the spectra are normalized to their intensity at high binding energy ($E_B = 22.5$ eV). The prominent feature in these spectra is the valence band, which is seen as a peak at about 3 eV binding energy. In addition one observes a broad structure around 12 eV binding energy. The intensity at the Fermi-energy ($E_F$) is very small. While the main valence band shows no strong dependence on the photon energy there, are clear changes in the states at 12 eV binding energy.

These changes around $E_B = 12$ eV are due to a resonance. To make them more visible, Fig. **3** shows difference spectra. The difference spectra were obtained by subtracting the spectrum taken with $h\nu = 72$ eV from each of the other spectra in Fig. **2**. By subtracting the spectrum taken at $h\nu = 72$ eV, before the onset of the resonance, one can remove the intensity coming from states other than the Cu-3d states.

The difference spectra show three features in the energy range between $E_F$ and $E_B = 15$ eV. The most prominent is again around $E_B = 3$ eV. These are the states comprising the valence band [16,17].

At higher binding energies there are two weaker features marked by dashed lines in Fig. **3**. One is a negative peak at about $E_B = 10.9$ eV. Its intensity grows with increasing photon energy and does not disappear for photon energies far above the threshold of $h\nu = 74.5$ eV. Because this feature does not disappear at high photon energies it is not clear whether or not this is a resonance effect or a cross section effect. The states at $E_B = 12.4$ eV on the other hand show a clear resonance. Such a resonance enhancement of Cu-states is well known for Cu-metal and its oxides. The resonance behavior of these so-called Cu-many body states was first studied by Thuler et al.[18]. Because of their complex resonance behavior even in Cu, CuO and $CuO_2$ they are not easily interpreted. It is however instructive to compare these many body states in the pure HTSC and in the insulating reference sample. To do so we have measured the similar energy distribution curves, shown in Fig. **2** for the pure material, for the insulating sample with 60% Pr-substitution. The difference spectra were obtained in the same way by subtracting the non-resonant spectrum at $h\nu = 72$ eV. The difference spectra are shown in Fig. **4**.

In this insulating sample, the dominant feature, in the Cu-partial density of states, is again the valence band around 3 eV binding energy. Comparing the pure HTSC in Fig. **3** to the insulating sample in Fig. **4** there are no significant changes in the partial Cu-density of states at the Fermi-level or in the valence band. In earlier investigations a shift of the total density of states of the valence band has been reported when going from the pure HTSC to the insulating compound [19][20]. This shift ($\Delta E \sim 0.5$ eV) is too small to be detected in the partial Cu-density of states shown here.

The Cu-features which were seen at 10.9 and 12.4 eV binding energy respectively in the pure sample, exhibit a different resonance response after Pr-substitution. While the feature at 10.9 eV has virtually disappeared in the insulating sample, the resonating Cu-many body state at $E_B = 12.4$ eV is almost unchanged. It shows a resonance comparable to that in the pure material.

The superconducting sample in Fig. **3** and the insulating sample in Fig. **4** show a very similar partial Cu-density of states at $E_F$ and in the valence band. Because these low-lying states, which dominate the superconducting properties, show no change when going from the superconductor to the insulator, we conclude that the states responsible for super-conductivity are predominantly O-states. This conclusion is supported by HREELS [21] measurements showing the O-2p character of the states at $E_F$. The virtual disappearance of the Cu-feature at 10.9 eV binding energy, however, indicates that there is some change in the Cu-states upon the Pr-substitution. The data suggest a coupling between electronic states of the Pr-ion and states on the Cu. This coupling is investigated in more detail in the next section.

**Partial Pr-4f density of states**

In this chapter the Pr-4f partial density of states is studied to obtain information about the electronic environment of the Pr-site in the insulating material. To do so we use the same resonant photoemission technique as in the preceding chapter. The only difference is that the photon energy is scanned across the binding energy of the Pr-4d core level at $E_B = 123.5$ eV [14]. Because in the Lanthanide series the 4f-level is partially filled it has a relatively low binding energy and contributes to the density of states in the region of the valence band. The important question is whether or not the rather localized 4f-states hybridize with other states, such as those in the $CuO_2$-plane.

Figure **5** shows the energy distribution curves in the region of the valence band for the insulating sample with 60% Pr-substitution for Ca. The spectra were taken at different photon energies from $h\nu = 120$ eV (top) to $h\nu = 135$ eV (bottom). While the majority of the valence band shows no change upon variation of the photon energy there is a clear resonant enhancement observable at a binding energy of $E_B = 1.3$ eV (dashed line). In the spectrum taken with $h\nu = 120$ eV (squares) the intensity at 1.3 eV binding energy is very low. When increasing the photon energy the intensity at 1.3 eV increases, reaches a maximum for $h\nu = 124$ eV (circles) and then decreases with further increasing photon energy.

The resonance effect observed here is much stronger than in the case of the Cu-resonances shown in the previous chapter (Figs. **2** and **3**). The reason is that the 4d-4f resonance is one of the so called Super-Coster-Kronig- or Giant-resonances. These are resonances where the incoming photon induces transitions to an intermediate state with same principle quantum number ($\Delta n = 0$) and difference in angular momentum of $\Delta \ell = 1$. These resonances are known to increase the photoemission signal by more than an order of magnitude[22].

The partial Pr-4f density of states is again obtained by subtracting the spectrum before the onset of the Pr4d-4f resonance ($h\nu = 120$ eV, squares) from that at resonance ($h\nu = 124$ eV, circles), $\Delta I(E) = I(124\text{eV}) - I(120\text{eV})$. It is illustrated in Fig. **6**. Aside from the subtraction no further normalization or other modification was performed. Because of the strong resonance enhancement the difference spectrum ΔI(E) should give an accurate estimate of the partial Pr-4f density of states.

The partial Pr-4f density of states consists of a single peak at about 1.3 eV binding energy. The shape of the peak is asymmetric. On the low binding energy side it shows a steep fall off and almost no intensity at the Fermi energy. Towards higher binding energy, on the other hand, it has a slowly decaying tail.

An asymmetric line shape is predicted for core levels interacting with a continuum of band states possessing a sharp cut-off at the Fermi-level. In this situation the creation of the core hole, by the incoming photon, leads to a screening response from the continuum states. As a result of this screening, electron-hole pairs are formed. The energy, necessary to form these electron-hole pairs, is taken from the outgoing electron, leading to a characteristic powerlaw fall off of the core level peak intensity towards higher binding energy. Such a lineshape was first calculated by Doniach and Šunjic[23]. The closed form solution for the Doniach-Šunjic lineshape is given by [24]

$$f(E) = \Gamma\{1-\alpha\} \frac{\cos\left(\alpha \frac{\pi}{2} + (1-\alpha)\arctan\left(\frac{E}{\gamma}\right)\right)}{\left(E^2 + \gamma^2\right)^{\left((1-\alpha)/2\right)}} \qquad (1)$$

Here $E = E_0 - E_B$ is the difference between the core level, situated at $E_0$, and the binding energy $E_B$, $\gamma$ is the width, $\alpha$ the dimensionless asymmetry parameter and $\Gamma$ the gamma-function which normalizes the spectral area under the curve. In an insulator with a band gap no low energy electron hole pairs can be formed. Therefore the Doniach-Šunjic line shape reduces to a symmetric Lorentzian with $\alpha=0$.

The solid line shown in Fig. **6** is a fit to equation (**1**) with a constant term added to simulate the background. Considering the signal to noise ratio the fit gives excellent agreement with the data. The parameters are $\alpha = (0.28\pm.05)$, $\gamma = (0.43\pm.08)$ eV and $E_0 = (1.36\pm.06)$ eV. The error margins given are conservative estimates from observing the fit quality.

In this case the asymmetry parameter $\alpha$ is the most important parameter. If this sample, which in Fig. **1** was shown to be insulating, had a real band gap, then the lineshape would be symmetric with $\alpha = 0$. This is not the case. While there are many parameters, intrinsic and extrinsic, influencing the width of core levels $\gamma$, none of them can explain an asymmetry. The asymmetry is a characteristic feature of the interaction between the core hole and the continuum of electronic states.

The non-zero asymmetry suggests that even for this insulating sample there is a finite density of states at $E_F$ in this insulating sample. Thus there can not be a genuine band gap. The asymmetry is however compatible with a mobility gap. In this case the electronic states at $E_F$ are spatially localized resulting in the characteristic variable range hopping observed in transport measurements [6, 8, 9, 10, 13]. As was shown by Chen and Kroha such localized states still give the asymmetric Doniach-Šunjic lineshape. In this case the spatial localization leads to a modification of the asymmetry parameter $\alpha$[25].

A similar asymmetric lineshape has been observed by Ratner et al. for metallic Pr-substituted $Bi_2Sr_2CuO_{6+y}$[26]. The low intensity of the Pr-4f state at $E_F$ lead these authors to the conclusion that the Pr-states do not hybridize strongly with the states at $E_F$, which are responsible for the transport properties and that Pr therefore does not affect superconductivity. The analysis in terms of a Doniach-Šunjic line shape which we have performed here, however, shows that there is substantial interaction between the Pr-4f states and the states at $E_F$. This seems to be the case for both the insulating Bi-2212 material studied here and the metallic Pr-substituted Bi-2201 studied by Ratner et al.[26]

The fact that this finite density of states at $E_F$ is observed at the site of the Pr-ion, located in between the $CuO_2$-bilayers, implies that even between the $CuO_2$-layers there is a finite density of states. Thus the electronic structure of Bi-2212 is not purely 2-dimensional. There must be a finite density of states between the $CuO_2$-layers leading to interlayer coupling. The experimental width of the Pr-4f level $\gamma = 0.43$ eV is consistent with the value of $\gamma = 0.40$ eV observed for metallic Pr[27].

After discussing the lineshape of the partial Pr-4f density of states we will, in the following, compare it to both Pr-metal and other Pr containing compounds. Resonant photoemission has been used on both pure Pr-metal [28] and metallic Pr-compounds [29, 30], to extract the partial Pr-4f density of states. In all these metallic Pr-compounds one observes a double peak structure for the Pr-4f density of states. These peaks are located at a binding energy of approximately 1.5 eV and 4.5 eV respectively. In these materials Pr is thought to be in the trivalent $4f^2$ configuration, which is consistent with observing two features in the partial 4f density of states. However, even Ce, having only one f-electron, exhibits two 4f-related peaks[31], which are well described by the Gunnarson-Schönhammer model [32].

The observation of only a single Pr-4f feature in these insulating cuprate compounds is therefore a puzzle. As shown in connection with Fig. **1** the electrical transport, the magnetic susceptibility[6, 7] and the Hall effect[7] give every reason to believe that in this compound Pr is also in the trivalent state. In addition the shape of the Pr4d-4f resonance, which will be shown in Fig. **7a** is very similar to that of metallic trivalent Pr and its compounds[29]. At the moment we have no explanation for this disagreement between the resistivity, susceptibility and Hall-effect data on one side and the partial Pr-4f density of states on the other side. One possible reason could be that the sample investigated here is an insulator, while all available data for partial Pr-4f density of states are for metallic materials. The decrease of the screening in this insulating compound could cause differences in the resonance response and could thus cause the disagreement between photoemission and transport data. In the literature there are no data available for the partial Pr-4f density of states in insulating materials to check this possibility.

**Coupling between Pr- and Cu-states**

In this section we will report the details of the resonance enhancement in the photoemission process at the Pr-4d threshold. This was done for initial states involving both Pr- and non Pr-levels. From comparing these resonances we conclude that there is significant coupling between the Pr-states and those in the $CuO_2$-plane.

The resonance enhancement in the photoemission process comes about because of the quantum mechanical interference of two different transitions leading from the ground state to the same final state[33]. This interference further enhances the effect due to the 4d → 4f electric dipole matrix element. For the 4f-4d resonance the states involved are well known[15]. The initial state $|i>$ is the groundstate $|\mathbf{Pr[4d^{10}, 4f^N]}>$ with a fully occupied 4d level and a 4f-level occupancy of N. The final state $|f>$ is that of a photoelectron $\varepsilon_k$ being ejected from the sample and a hole in the Pr-4f level, $|\mathbf{Pr[4d^{10}, 4f^{N-1}], \varepsilon_k}>$. The intermediate state $|m>$ is an unstable, excited state with a hole in the Pr-4d level and an extra electron in the Pr-4f level, $|\mathbf{Pr[4d^9, 4f^{N+1}]}>$. While the intermediate state $|m>$ has a well defined energy and is thus discrete, the photoelectron $\varepsilon_k$ can have any value of kinetic energies and is thus in a continuum of states. The problem is therefore similar, to the general problem first discussed by Fano[34], of the interaction between a continuum state and a discrete state. In fact, it can be shown that the Fano line shape is a special case of this more general form[33].

The two possible transitions between the initial and the final state are[33] the direct transition

$$|Pr[4d^{10}, 4f^N]> \quad\Rightarrow\quad |Pr[4d^{10}, 4f^{N-1}], \varepsilon_k> \tag{2}$$

and the indirect transition,

$$|Pr[4d^{10}, 4f^N]> \Rightarrow |Pr[4d^9, 4f^{N+1}]> \Rightarrow |Pr[4d^{10}, 4f^{N-1}], \varepsilon_k> \tag{3}.$$

Both, the transition to the intermediate state $|m>$ and the direct transition are electrical dipole transitions described by the dipole operator $T_D$. The intermediate state on the other hand decays through an Auger process. This is described by the Auger operator $T_A$[33]. Because of the interference between the two transitions the total cross section is given by the square of the sum of the transition amplitudes for the direct and the indirect transition, $\tau_{ind}$ and $\tau_{dir}$ respectively,

$$\sigma(E) = |\tau_{dir} + \tau_{ind}(E)|^2. \tag{5}$$

The indirect transition has a strong energy dependence because of the threshold for exciting the intermediate state $(E = h\nu > E_B(Pr-4d))$. The direct transition on the other hand has a negligible energy dependence in the energy region of interest. Using Fermi's golden rule for the transition amplitudes this gives[33]

$$\sigma(E) = \left| \langle f|T_D|i\rangle + \frac{\langle f|T_A|m\rangle\langle m|T_D|i\rangle}{(E - E_0 + i\Gamma/2)} \right|^2 \quad (4)$$

Here |m> denotes the intermediate state which in our case would be the excited Pr[$4d^9\,4f^{N+1}$] configuration. We will assume it to be the only intermediate state. $E_0$ is the binding energy of the core level, renormalized by the interaction with the other states, and $\Gamma$ is the lifetime of the excited state. Equation (**4**) leads to a resonance, which in the most general case is a combination of a Lorentzian and a Fano lineshape[33]. For simplicity we will assume it to be a pure Fano resonance which is described by

$$\sigma(\varepsilon) = (\varepsilon+q)^2/(1+\varepsilon^2) \quad (6).$$

Here $\varepsilon = (E-E_0)/\Gamma$ is a normalized energy scale. The lifetime of the Auger state $\Gamma$ is given by[33]

$$\Gamma = \pi\,|<f|T_A|m>|^2. \quad (7).$$

The parameter q in equation (**6**) determines the shape of the resonance. The square of this parameter is the ratio of transitions to the intermediate excited state |m> relative to the direct transitions into the continuum of final states[33],

$$q^2 = \frac{1}{\pi \cdot \Gamma} \left| \frac{\langle m|T_D|i\rangle}{\langle f|T_D|i\rangle} \right|^2 \quad (8).$$

The energy dependent corrections $\sigma(E)$ for various initial states |i> are shown in Fig. **7**. The spectra were collected in the constant initial state (CIS) mode. Here the photon energy and the kinetic energy of the detected electrons is swept simultaneously so that electrons from the same initial state |i>, $E_i$ are detected; $E_{kin} = h\nu - \phi - E_i$. Where $\phi$ is the work function of the system, which is $\phi = 4.45$ eV in our case. All spectra are normalized

to the photon flux and their intensity at $h\nu = 115$ eV. Their amplitude thus gives information about the magnitude of the resonance effect. The spectra are offset vertically for clarity. Panel (**a**) shows the CIS-spectrum for the Pr-4f state at $E_i = 1.45$ eV. The strong resonant enhancement at $h\nu = 124$ eV caused by the Super-Coster-Kronig resonance is seen as a 12-fold increase of intensity. It reflects the growth of the 4f-state already seen in Fig. **5**. The solid line is a fit to a Fano resonance. To take the energy dependence of the direct transition into account a linear background was added to equation (**6**),

$$\sigma(E) = a + b \cdot E + \sigma_r \cdot \frac{(q+\varepsilon)^2}{(1+\varepsilon^2)} \qquad (9).$$

The parameter $\sigma_r$ is a measure of the strength of the resonant enhancement. All fit parameters are given in table (**1**). The resonances are well described by the Fano line shape. Deviations only occur before and after the peak region. Before the onset of the resonance the pure Fano lineshape has a minimum which is not observed in this case. This region could be fit better by using the general combination of a Fano- and a Lorentzian-resonance in equation **5** [33]. However this would add an additional parameter and is not necessary for the following discussion. Above the resonance the experimental data show an additional shoulder, which can not be reproduced by the Fano line shape. This shoulder is also seen in pure Pr and its metallic compounds[29]. To minimize the influence of this shoulder on the fit parameters, the fit was limited to the region shown by the solid line, which excludes the shoulder.

The fit parameters for the Pr-4f initial state are $q = 1.9$, $\Gamma = 3.6$ eV and $E_0 = 123.2$ . They are also given in table **1** together with those for other initial states. Comparing the resonance to that of the metallic $PrCo_2$ for example we find that the shape is very similar [29]. While for this compound there are no numerical values available we can compare the resonance to that in Yb[35] and Ce[36] for example. Here one finds $q = 2.4 / 2.4$ and $\Gamma = 1.7 / 1.3$ eV respectively. These are similar values which gives us confidence in our analysis.

To investigate the possibility of coupling between Pr-states and states in the $CuO_2$-plane, which is the main objective of this paper we have recorded similar CIS spectra for a number of other initial states. These spectra are illustrated in Fig. **7b**. From top to bottom the initial states are the Cu-feature at $E_i = 10.0$, The Cu-many body state at $E_i = 12.75$ eV, a state just below the Fermi-level $E_i = 0.25$ eV and the Bi-5d core level at $E_i = 28.35$ eV. For comparison we also show a constant final state scan (CFS) for the same sample and energy range.

In the constant final state mode the photon energy is varied, while the kinetic energy of the photoelectron, the final state, is kept constant. By using a kinetic energy in the range of the inelastically scattered secondary electrons, one obtains a signal which is proportional to the total cross section for photoemission. Because the overwhelming fraction of photoemission processes is non resonant this signal should give a good estimate of the non resonant cross section. Any difference between the CFS and CIS spectra are thus caused by resonant processes. In the CFS spectrum, the Pr-4d core level should show up as a step. This is not observed and indicates a very small crossection for direct photoemission from the Pr-4d level.

Again the spectra are normalized to the photon flux and to the intensity at $h\nu = 115$ eV and are offset vertically. Both the Cu features at 10.0 and 12.75 eV binding energy and the spectrum taken with an initial state close to $E_F$ show a pronounced structure at the Pr 4d-4f resonance. The Bi-core level on the other hand shows no structure. The resonance enhancement of the photoemission cross section from Cu features when scanning the photon energy across the Pr-4d core level shows that the Pr- and Cu-states couple. The Bi-state on the other hand seems to have no coupling because it shows no structure at this photon energy. As in part (**a**) of Fig. **7** the solid line is again a fit to a Fano resonance (equation **9**). The fit gives good results for the two Cu features and for $E_i = 0.25$ eV. The spectrum of the Bi-core level shows a very slight enhancement at ~120eV but this is weaker by more than an order of magnitude and can not be described by a Fano resonance.

Table (**1**) gives the fit parameter for the four initial states showing resonances with the Pr-4d core level together with conservative estimates of the error bars. These are obtained by observing the fit quality. The range of the fits is equal to the region for which the solid fit line is shown in Fig. **7**. As discussed above, the Pr-4f state has by far the strongest resonance expressed by the 10 times higher $\sigma_r$ value. The shape value q and the width $\Gamma$ are comparable to those of other rare earth materials[35, 36]. This gives us confidence in our interpretation of the resonance process and the states involved in it. The fit parameters show a number of similarities and trends. It is significant that all initial states |i> have the same shape parameter q = 1.8 and the same resonance energy $E_0$ = 123 eV within our error bars. This supports our assumption that the excited intermediate state is always the same. As expected the intraatomic resonance of the Pr-4f state is the strongest. The resonances of the other initial states are weaker by an order of magnitude. While the initial state close to $E_F$ at $E_i$ = 0.25eV could be argued too be just the foot of the Pr-4f state at $E_i$ = 1.45 eV, the Cu features are much to far away to still be influenced by the narrow Pr-4f state. With binding energies 8.5 eV and 11.5 eV below that of the Pr-4f state, which has a width of $\gamma$ = 0.43 eV (Fig. **5**), these resonances are certainly independent features. The fact that both the Cu features and the state just below $E_F$ show a Fano-resonance at the energy of the Pr-4d core level points to a coupling between Pr-states and these initial states. Given the resonance process this coupling indicates an Auger decay of the intermediate Pr[$4f^{N+1}$]-state into a state with a hole in the respective initial state. At present the details of this Auger transition are not clear. Because of the very small, almost atomic size of the involved orbitals a direct overlap of the Pr[$4f^{N+1}$] wave function with the Cu orbitals seems unlikely. An indirect transition involving the still somewhat extended states at $E_F$ is more likely. The observed possibility of different decay channels for the intermediate state and the widths of the Fano resonances indicates a very short lifetime $\Delta t$ of this excited Pr[$4d^9\ 4f^{N+1}$] state. It can be estimated by the uncertainty relation $\Delta t = \Delta E/\hbar$. Because there are different decay channels the energy uncertainty is given by the sum of all the uncertainties or the sum of the fitted $\Gamma$ values. This gives an estimated lifetime of $\Delta t \sim 10^{-17}$ s. A puzzling feature is that the Auger decay to the Pr-4f configuration has the smallest $\Gamma$ and thus the largest lifetime. All other decay channels have faster relaxation rates. The very rapid decay of the intermediate state on the other

hand may be an explanation why the Pr-4d core level is almost unobservable in the regular energy distribution curves and is also absent in the CFS spectrum of figure **7b**. Such a situation is for example observed in Xe [33, 37]. Because the Cu features and the states just below $E_F$ show a resonance when scanning the photon energy across the Pr-4d core level we conclude that there is a finite coupling between Pr-states and those in the $CuO_2$-plane. This coupling seems to be absent for Bi-states, which are spatially removed and where we do not observe a Fano-resonance. Because there is a finite coupling from the Pr-states to the Cu-states, which are involved in the conduction process, the conducting states experience a contribution to their electrostatic potential, originating from the randomly distributed Pr-ions. This random potential fluctuation can lead to spatial localization of the electronic states and thus to the metal-insulator transition which is experimentally observed in these materials.

**Conclusion**

Substitution of Pr for Ca induces a metal-insulator transition in Bi-2212. Using resonant photoemission we have shown that there is substantial coupling between the Pr-states and states in the $CuO_2$-plane. This coupling is seen as a Fano resonance observed for the Cu-many body states and the states at $E_F$ when scanning the photon energy across the binding energy of the Pr-4d core level. In the insulating material the Pr-4f state shows a Doniach-Šunjic line shape, which is characteristic of the interaction with a continuum of electronic states with a sharp cutoff at $E_F$. This shows that the metal-insulator transition is not caused by the opening of a band gap. Rather it supports the view of a mobility gap caused by spatial localization of the carriers. One origin of the spatial localization could be the random electrostatic potential seen by the states in the $CuO_2$-plane because of their coupling to the statistically distributed Pr-ions.

Our results show that in order to better understand HTSC one needs to consider the coupling of the $CuO_2$-plane to both, the intermediate Ca/**RE**-O layer and the next $CuO_2$-layer. In addition to this the role of disorder in these non stoichiometric materials will have to be studied in detail.

**Acknowledgments**

We would like to thank D. Quitmann and H. Kroha for helpful conversations. The help of Jian Ma in setting up the UHV-system and the technical assistance of Shawn Heldebrandt are greatly appreciated. Financial support for this project was provided by the Deutsche Forschungsgemeinschaft, the National Science Foundation, both directly and through support of the Synchrotron Radiation Center and by the Bundesministerium für Forschung und Technologie.

**Figure captions:**

**Figure 1 :** Electrical resistivity vs. temperature for $Bi_2Sr_2[Ca_{1-x}Pr_x]Cu_2O_{8+y}$. Note the transition from metallic ($d\rho/dT > 0$) to insulating ($d\rho/dT < 0$) behavior at $x_C = (0.49\pm0.02)$. The samples on which photoemission spectra are reported are samples with x=0 and x=0.6 respectively.

**Figure 2 :** Photoemission spectra for pure $Bi_2Sr_2CaCu_2O_{8+y}$ HTSC with photon energies in the range of the Cu-3p core level binding energy at 74.5 eV. The spectra are normalized to their intensity at $E_B = 22.5$ eV and offset vertically for clarity.

**Figure 3 :** Partial Cu-density of states for HTSC $Bi_2Sr_2CaCu_2O_{8+y}$. The spectra are obtained from those in Fig. **2** by subtracting the spectrum at $h\nu = 72$ eV, before the onset of the Cu-resonance, from the spectra at a given photon energy ($h\nu$).

**Figure 4 :** Partial Cu-density of states for insulating $Bi_2Sr_2[Ca_{0.4}Pr_{0.6}]Cu_2O_{8+y}$. The spectra are obtained by subtracting the spectrum at $h\nu = 72$ eV, before the onset of the Cu-resonance, from the spectra at a given photon energy ($h\nu$).

**Figure 5 :** Valence band spectra of an insulating $Bi_2Sr_2[Ca_{0.4}Pr_{0.6}]Cu_2O_{8+y}$ sample at different photon energies in the range of the Pr-4d core level binding energy. Note the resonant enhancement of the Pr-4f state at $E_B = 1.3$ eV (dashed line) for $h\nu = 124$eV. The spectra before and at resonance are marked by squares and circles respectively.

**Figure 6 :**  Partial Pr-4f density of states for an insulating $Bi_2Sr_2[Ca_{0.4}Pr_{0.6}]Cu_2O_{8+y}$ sample. The partial Pr-4f density of states is obtained by subtracting the spectrum before the onset of the Pr 4d-4f resonance (hν = 120 eV) from the spectrum at resonance (hν = 124 eV) in Figure **5**. The solid line is a fit to a Doniach-Šunjic lineshape with free parameters γ (width), $E_0$ (position) and α (asymmetry). They are discussed in the text.

**Figure 7 :**  Constant initial state spectra for an insulating $Bi_2Sr_2[Ca_{0.4}Pr_{0.6}]Cu_2O_{8+y}$ sample. Part (**a**) shows the Super-Coster-Kronig Pr 4p-4f resonance, part (**b**) shows spectra for other initial states originating from the $CuO_2$-plane, the Bi-5d core level and a constant final state spectrum as a reference spectrum. The solid lines are fits to a pure Fano resonance (equation **9**).

**Table 1:**

Fit parameters for Fano lineshape of the resonant photoemission enhancement observed at the threshold of the Pr-4d core level binding energy at $E_B$ = 123.5 eV. The parameters correspond to equation **9** and are explained in the text.

| State | $E_i$[eV] | q | Γ [eV] | $E_0$ [eV] | $σ_r$ | A | B |
|---|---|---|---|---|---|---|---|
| $E_F$ | 0.25 | (1.8±.2) | (3.6±.5) | (122.3±.3) | (0.22±.01) | 2.8 | 0 |
| Pr-4f | 1.45 | (1.9±.2) | (2.4±.2) | (123.2±.2) | (2.3±.3) | (1.0±.1) | 0 |
| Cu-feature | 10.1 | (1.7±.1) | (6.0±.2) | (123.0±.2) | (0.21±.01) | (1.3±.1) | (.024±.001) |
| Cu-many body | 12.75 | (1.9±.1) | (7.3±.3) | (123.3±.2) | (0.22±.02) | (1.5±.2) | (.030±.001) |

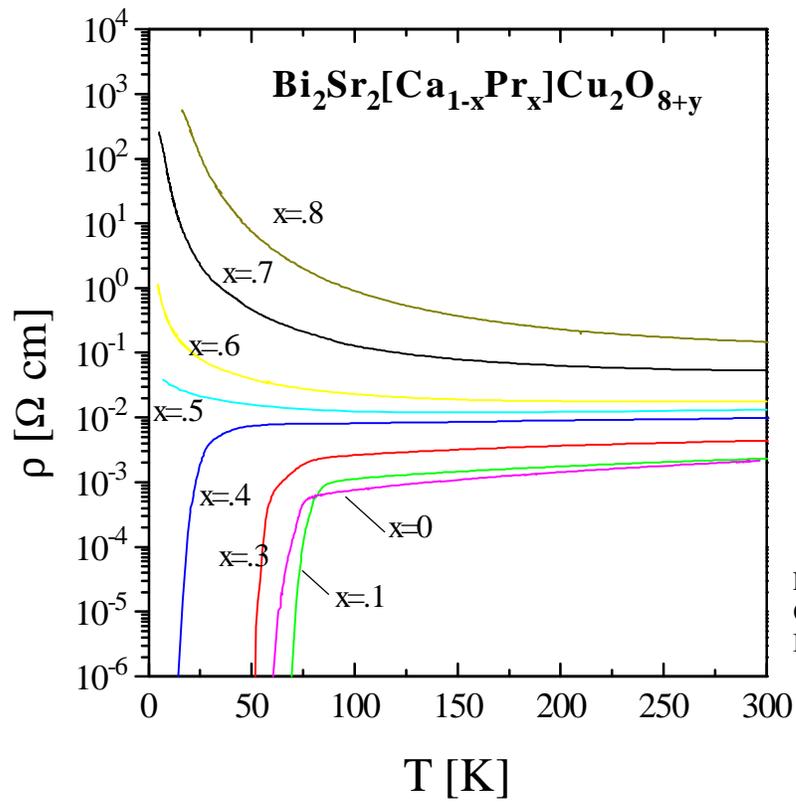

Fig. 1:
Quitmann et al.
Interlayercoupling and ...

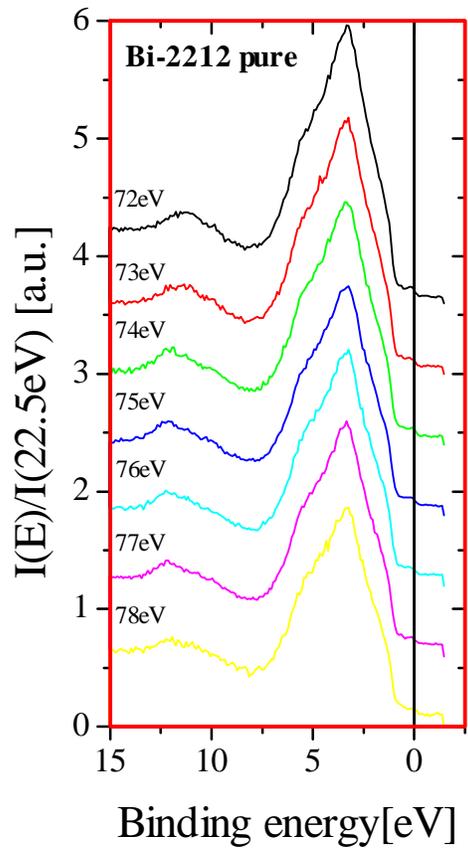

Fig. 2
Quitmann et al.
Interlayer coupling and•

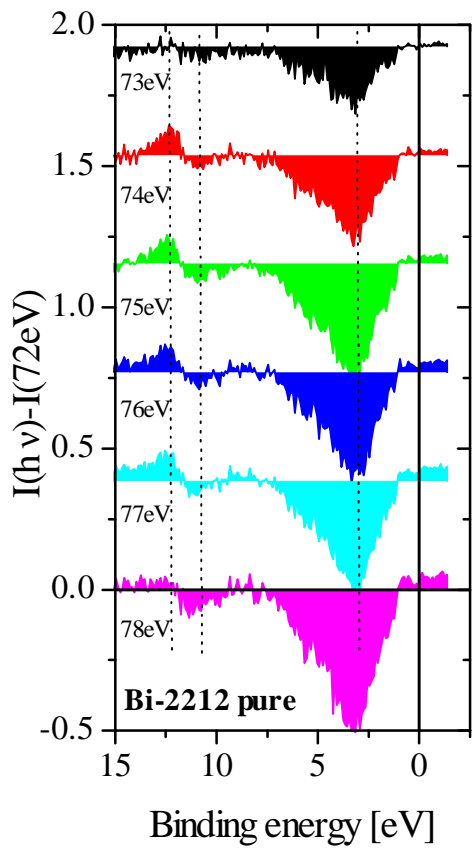

Fig. 3
Quitmann et al.
Interlayer coupling and•

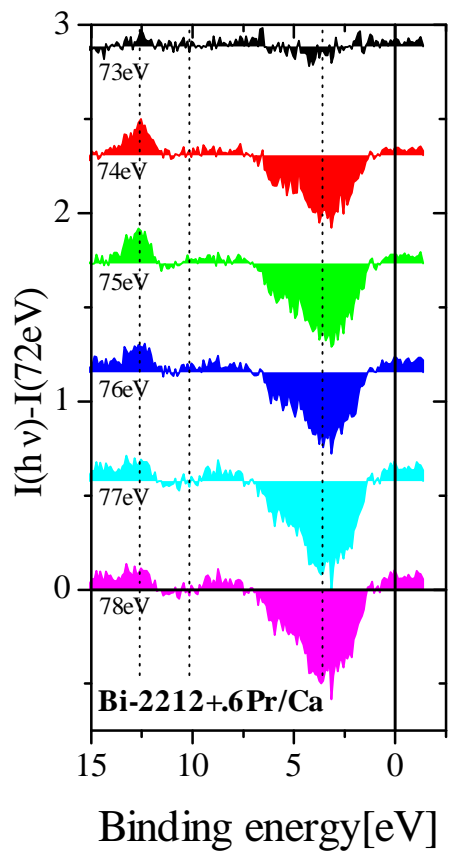

Fig. 4
Quitmann et al.
Interlayer coupling and•

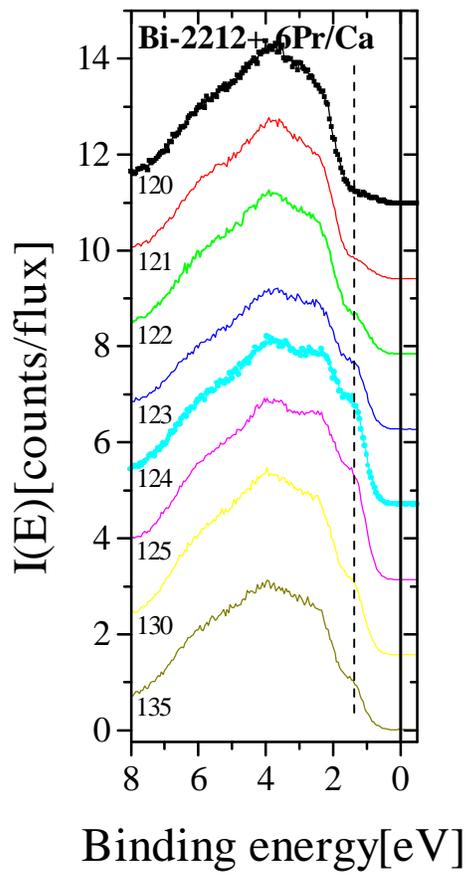

Fig. 5
Quitmann et al.
Interlayer coupling and•

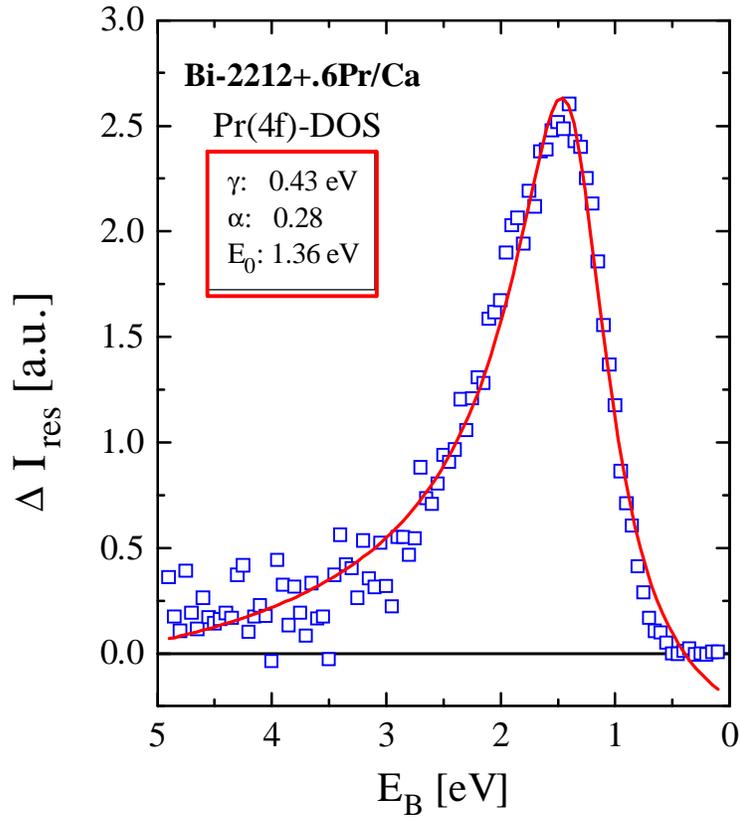

Fig. 6
Quitmann et al.
Interlayer coupling and

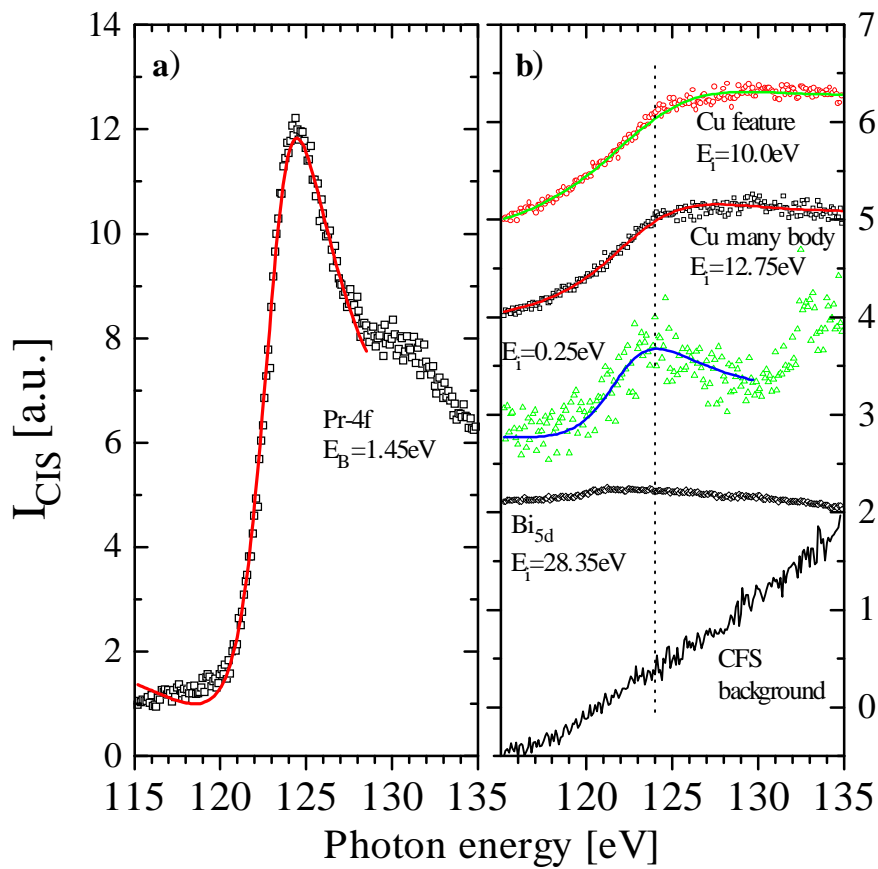

Fig. 7 Quitmann et al. Interlayer coup.

Quitmann et al.		Interlayer Coupling ...		- 29 -